\documentclass[useAMS,usenatbib]{mn2e}

\usepackage{amssymb}
\usepackage{graphicx}

\def\lap{\hbox{${_{\displaystyle<}\atop^{\displaystyle\sim}}$}}
\def\gap{\hbox{${_{\displaystyle>}\atop^{\displaystyle\sim}}$}}

\title[High-energy neutrinos from pulsars]
{Flux predictions of high-energy neutrinos from pulsars}
\author[Bennett Link and Fiorella Burgio]
{Bennett Link$^{1}$\thanks{E-mail:link@physics.montana.edu}
and Fiorella Burgio$^{2}$\thanks{E-mail:fiorella.burgio@ct.infn.it}\\
$^{1}$Department of Physics, Montana State University, Bozeman,
Montana 59717, USA \\
$^{2}$INFN Sezione di Catania, Via S. Sofia 64, I-95123 Catania, Italy}

\begin{document}

\date{Accepted 2006 June 06. Received 2006 June 05; in original form 2006 April 12 }

\pagerange{\pageref{firstpage}--\pageref{lastpage}} \pubyear{2005}

\maketitle

\label{firstpage}

\begin{abstract}
Young, rapidly rotating neutron stars could accelerate ions from their
surfaces to energies of $\sim 1$ PeV. If protons reach such energies,
they will produce pions (with low probability) through resonant
scattering with x-rays from the stellar surface. The pions
subsequently decay to produce muon neutrinos. Here we calculate the
energy spectrum of muon neutrinos, and estimate the event rates at
Earth. The spectrum consists of a sharp rise at $\sim 50$ TeV,
corresponding to the onset of the resonance, above which the flux
drops with neutrino energy as $\epsilon_\nu^{-2}$ up to an
upper-energy cut-off that is determined by either kinematics or by the
maximum energy to which protons are accelerated. We estimate event
rates as high as 10-100 km$^{-2}$ yr$^{-1}$ from some candidates, a
flux that would be easily detected by IceCube. Lack of detection would
allow constraints on the energetics of the poorly-understood pulsar
magnetosphere.

\end{abstract}


\section{Introduction}

Astrophysical neutrinos of high energy ($\gap 1$ GeV) are expected to
arise in many environments in which protons are accelerated to
relativistic energies.  Neutrinos produced by the decay of pions
created through hadronic interactions ($pp$) or photomeson production
($p\gamma$) escape from the source and travel unimpeded to Earth, and
carry information directly from the acceleration site. Neutrinos may
be produced by cosmic accelerators, like those in supernova remnants
\citep{pro98}, active galactic nuclei
\citep{lm00}, micro-quasars \citep{dist02} and gamma-ray bursts
\citep{wb97,dai01}. To detect these neutrinos, several projects 
are underway to develop large-scale
neutrino detectors under water or ice; AMANDA-II, ANTARES and Baikal are
running, while IceCube \citep{ha06}, NEMO and NESTOR
\citep{ca03} are under construction.

Recently, we proposed that young ($t_{age}\lesssim 10^5
\rm yr$) and rapidly-rotating neutron stars could be  intense
neutrino sources (Link \& Burgio 2005; hereafter LB05). For stars that
have a stellar magnetic moment with a component anti-parallel to the
spin axis (as we expect in half of neutron stars), ions will be
accelerated off of the surface; otherwise, electrons will be
accelerated. If energies of $\sim$ 1 PeV per proton are attained,
pions will be produced through photomeson production as the protons
scatter with surface x-rays, producing a beam of $\mu$ neutrinos with
energies above $\sim 50$ TeV. Detection of such neutrinos would
provide an invaluable probe of the particle acceleration processes that
take place in the lower magnetosphere of a neutron star. In this
paper we predict the neutrino spectrum that would result from this
production mechanism to aid in the interpretation of experimental
results from searches for astrophysical neutrinos. We obtain improved
estimates of the event rate.

In the next section, we review the acceleration model of LB05. In
Section III, we calculate the neutrino spectrum that results from this
model. In Section IV, we estimate the count rates that could be seen
in a km$^2$-scale experiment such as IceCube. We conclude with a
discussion of the prospects for detection of high-energy neutrinos
from pulsars.

\section[]{The Model}

In the neutrino production scenario of LB05, protons (within or
without nuclei) are accelerated in the neutron star magnetosphere to
high enough energies to undergo resonant scattering with surface x-ray
photons (the $\Delta$ resonance):
\begin{equation}
p\gamma\rightarrow \Delta^+\rightarrow n\pi^+ \rightarrow
n\nu_\mu\mu^+ \rightarrow n \nu_\mu e^+ \nu_e\bar{\nu}_\mu.
\label{reaction}
\end{equation}
The $\Delta^+$ is a short-lived excited state of the proton with a
mass of 1232 MeV.  For this process to be effective, ions must be
accelerated close to the stellar surface, where the photon density is
high and the process is kinematically allowed (see below, and
LB05). The plasma is tied to the magnetic field, so the acceleration
can occur only in the direction of the magnetic field ${\mathbf B}$.
In a quasi-static magnetosphere with a magnetic axis that is parallel
or anti-parallel to the rotation axis, the potential drop {\em across}
the field lines of a star rotating at angular velocity $\Omega=2\pi/p$
(where $p$ is the period) from the magnetic pole to the last field
line that opens to infinity is of magnitude \citep{gj69}:
\begin{equation}
\Delta\Phi=\frac{\Omega^2 B R^3}{2c^2}\simeq 7\times 10^{18}
B_{12} R_6^3 p_{\rm ms}^{-2} \mbox{ Volts} \label{phimax}, 
\end{equation}
per ion of mass number $A$ and charge $Z$. 
Here $B=10^{12}B_{12}\rm G$ is the strength of the dipole component
of the field at the magnetic poles, $R=10^{6}R_{6}$ cm is the
stellar radius and $p_{\rm ms}$ is the spin period in
milliseconds. Henceforth we take $R_6=1$ when making estimates. 
In equilibrium (not realized in a pulsar), a co-rotating
magnetosphere would exist in the regions above the star in which
magnetic field lines close; the charge density would be $\rho_q
\simeq eZn_0\simeq B/pc$ (cgs), where $n_0$ is the Goldreich-Julian
number density of ions. Deviation from corotation will lead to
charge-depleted gaps somewhere above the stellar surface, through
which charges will be accelerated to relativistic energies
\citep{rs75,as79}. 
The proton energy threshold $\epsilon_p$ for $\Delta^+$ production
is given by
\begin{equation}
\epsilon_p\epsilon_\gamma\ge (1-\cos\theta)^{-1}\ 0.3\ {\rm GeV}^2, 
\end{equation}
where $\epsilon_\gamma$ is the photon energy and $\theta$ is the
incidence angle between the proton and the photon in the lab frame.
Young neutron stars typically have temperatures of $T_\infty \simeq
0.1 \rm \ keV$, and photon energies $\epsilon_\gamma=2.8
kT_\infty(1+z_g)\sim 0.4$ keV, where $z_g\simeq 0.4$ is the
gravitational red-shift (for $R=9$ km and $M=1.4M_\odot$) and
$T_\infty$ is the surface temperature measured at infinity. For
$\theta\simeq\pi/2$, corresponding to scattering of protons near the
stellar surface with photons from the stellar horizon, the proton
threshold energy for the $\Delta$ resonance is then $\epsilon_{p,{\rm
th}}\simeq T_{\rm 0.1keV}^{-1}$ PeV, where $T_{\rm 0.1keV}\equiv
(kT_\infty/0.1\mbox{ keV})$. Let us compare the required energy for
resonance to the potential drop per proton {\em across} ${\mathbf B}$:
\begin{equation}
\frac{\epsilon_{p,{\rm th}}}{\Delta\Phi/A} \sim 
10^{-4} B_{12}^{-1}p_{\rm ms}^2 T_{\rm 0.1keV}^{-1} A. 
\end{equation}
Hence, for a pulsar spinning at 10 ms, a potential drop {\em along}
${\mathbf B}$ of only $\sim 1$\% of that {\em across} ${\mathbf B}$ will be
sufficient to bring protons or low-mass nuclei to the
resonance. Optimistically assuming that the full potential
$\Delta\Phi$ is available for acceleration along field lines, a
necessary condition for the $\Delta$ resonance to be reached is ($A=1$) 
\begin{equation}
B_{12} p_{\rm ms}^{-2} T_{\rm 0.1keV} \ge 3\times 10^{-4}.
\label{threshold}
\end{equation}
Assuming $T_{\rm 0.1keV}=1$, typical of pulsars younger than $\sim
10^5$ yr, there are 10 known pulsars within a distance of 8 kpc that
satisfy this condition (Manchester et al. 2005), 
about half of which should have
positively-charged magnetic poles; these are potentially detectable
sources of $\mu$ neutrinos. The best
candidates are young neutron stars, which are usually rapidly spinning
and hot. Equality corresponds to the full potential across field lines
being present along field lines, probably an unlikely scenario, since
space charge will act to quench the electric field along ${\mathbf
B}$. For stars in which the above inequality is easily satisfied, the
protons will reach energies sufficient to undergo photomeson
production if the electric field along a typical open field line is
much smaller than the field across the line, the more probable
situation.

In the photomeson production process of eq. (\ref{reaction}), each
muon neutrino receives 5\% of the energy of the proton. [We assume,
for simplicity, that the pions do not undergo subsequent
acceleration]. Typical proton energies required to reach resonance are
$\sim T_{\rm 0.1keV}^{-1}$ PeV, so the expected $\mu$ neutrino
energies will be $\epsilon_\nu\sim 50 T_{\rm 0.1keV}^{-1}$
TeV. Moreover, since the accelerated protons are far more energetic
than the radiation field with which they interact, any pions produced
through the $\Delta$ resonance, and hence, any muon neutrinos, will be
moving in nearly the same direction of the protons. The radio and
neutrino beams should be roughly coincident, so that some radio
pulsars might also be detected as neutrino sources. We see the radio
beam for only a fraction $f_b$ of the pulse period, the {\em duty
cycle}. Typically, $f_b\simeq 0.1-0.3$ for younger pulsars.  We take
the duty cycle of the neutrino beam to be $f_b$ (but see below). In
LB05, we estimated the phase-averaged neutrino flux at Earth resulting
from the acceleration of positive ions, at a distance $d$ from the
source, to be
\begin{equation}
\phi_\nu\simeq c f_b f_d n_0 \left(\frac{R}{d}\right)^2
P_c, 
\end{equation}
where $f_d$ is the fraction by which the space charge in the
acceleration region is depleted below the corotation density $n_0$,
and $P_c$ is the conversion probability. A better estimate is
\begin{equation}
\phi_\nu\simeq 2c f_b f_d(1-f_d) n_0 \left(\frac{R}{d}\right)^2
P_c, 
\label{nuflux}
\end{equation}
The pre-factor
$f_d(1-f_d)$ is a more realistic interpolation between the two regimes
of complete depletion ($f_d=0$) and no depletion ($f_d=1$). In the
latter case, there should be no neutrino production as the field along
${\mathbf B}$ is entirely quenched. The factor of two arises from
inclusion of the $\bar\nu_\mu$ from eq. (\ref{reaction}). 
For the purposes of
calculating the spectrum, we use the following differential form for
the 
neutrino flux: 
\begin{equation}
\frac{d\phi_\nu}{d\epsilon_\nu}=2cf_bf_d(1-f_d)n_0\left(\frac{R}{d}\right)^2
\frac{dP_c}{d\epsilon_\nu}. 
\label{diffflux}
\end{equation}
Here we are interested in estimating upper limits on the
flux; we henceforth take $f_d=1/2$ and $Z=A=1$.
 
In LB05, we estimated $P_c$ by assuming that the protons reach
the energy required for photomeson production immediately above the
stellar surface; we then evaluated $P_c$ at a particular height above
the stellar surface. The focus of this paper is to obtain the spectrum
$dP_c/d\epsilon_\nu$. We account for the fact that the proton
acceleration will take place over a finite distance above the
surface. We also include the finite width of the $\Delta$ resonance in
the cross section. 

\section[]{Neutrino Spectrum}

Let the accelerated proton have energy $\epsilon_p$ and the surface
photons have energy $\epsilon_\gamma$. Define a dimensionless energy
$x=\epsilon_p\epsilon_\gamma/\epsilon_0^2$, where $\epsilon_0^2\equiv
0.3$ GeV$^2$. To account for the finite width of the $\Delta$
resonance, let us express the resonant contribution to the
energy-dependent $p\gamma$ cross section for a proton of energy
$\epsilon$ in the nucleon rest frame as
\begin{equation}
\sigma=\sigma_0\exp\left(\frac{-(\epsilon-\epsilon_T)^2}{2w^2}\right)
\qquad \mbox{for $\epsilon>\epsilon_T-w$},
\end{equation}
where $\sigma_0=5\times 10^{-28}$ cm$^{-2}$ is the cross section for
$\Delta^+$ production, $\epsilon_T$ is the $\Delta$ energy threshold
and $w\simeq 100$ MeV is the width of the $\Delta$ resonance. For a
given photon energy and scattering angle, the proton energy at the
resonance is given by
\begin{equation}
\epsilon_T=\frac{\epsilon_0^2}{\epsilon_\gamma}(1-\cos\theta)^{-1}
\Longrightarrow 
x_T=(1-\cos\theta)^{-1}.
\end{equation}
The cross-section is non-zero for $x>x_T-x_w$ where $x_w\equiv
w\epsilon_\gamma/\epsilon_0^2\simeq 0.1$. This implies a minimum
kinematic scattering angle $\theta_0$ given by
\begin{equation}
\cos\theta_0=1-(x+x_w)^{-1}.
\end{equation}
\begin{figure}
\centering
 \includegraphics[width=6cm]{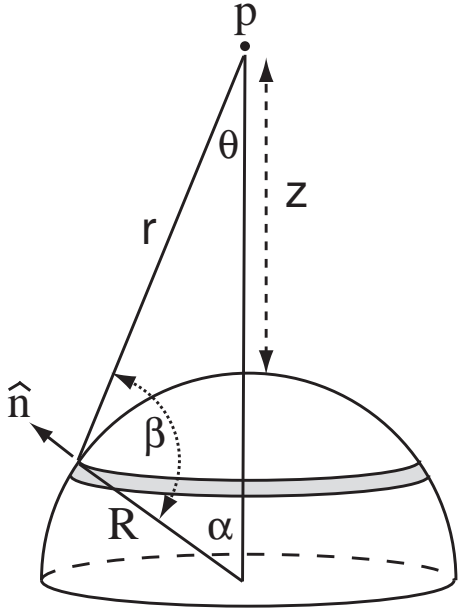}
  \caption{Geometrical setup. See text for details.}\label{geom}
\end{figure}
In Fig. 1 we show the geometry that we use for calculating the
spectrum. We neglect the effects of gravitational light bending, which
would act to increase the rates we calculate here. The proton is
located at the point denoted by $p$, at a height $z$ above the stellar
surface, and $\theta$ is the scattering angle between the proton and
the incoming photon. The scattering angle $\theta$ is always less than
$\pi/2$, which requires the protons to have energies that
satisfy $x>1-x_w$ in order to undergo resonant scattering. There is
also a maximum scattering angle $\theta_h$ defined by the stellar
horizon as seen from height $z$:
\begin{equation}
\sin\theta_h=(1+z)^{-1}, \label{horizon}
\end{equation}
where $z$ is the height above the stellar surface in units of the
stellar radius $R$. We must have $\theta_0<\theta_h$ for resonance to occur.
 We can rewrite
the cross-section as
\begin{equation}
\sigma(x,\theta)=\sigma_0\exp\left(\frac{-(x-x_T)^2}{2x_w^2}\right)
\qquad \mbox{for $x>x_T-x_w$}.
\end{equation}

\begin{figure*}
\centering
 \includegraphics[width=5cm, angle=270]{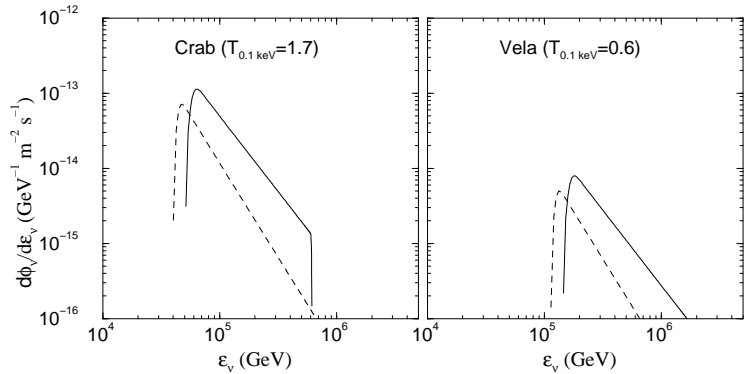}
  \caption{The neutrino energy flux is displayed for the Crab and the Vela pulsars,
  for the cases of linear (solid line) and quadratic proton acceleration
  (dashed line). }\label{spectrum}
\end{figure*}
\noindent
To calculate the conversion probability, we need the
number density of photons as a function of height $z$ and incident
photon angle $\theta$. Let the number density of photons at the
stellar surface be $n_\gamma(0)$. An element of surface radiates into $2\pi$
steradians, independent of angle. For a small surface element of
area $dA$, the contribution $dn$ to the total photon number density
$n$ is
\begin{equation}
dn=\frac{n_\gamma(0)}{2\pi}\frac{dA}{r^2}\,\hat{n}\cdot\hat{r}\qquad\qquad
\hat{n}\cdot\hat{r}=\cos(\alpha+\theta).
\end{equation}
Taking the ring surface element shown in the figure, that area is
\begin{equation}
dA=2\pi R^2\sin\alpha\, d\alpha.
\end{equation}
Some useful trigonometric identities are
\begin{equation}
r\sin\theta=R\sin\alpha \qquad (1+z)\sin\theta=\sin\beta \qquad
\beta =\pi - \alpha - \theta,
\end{equation}
which give
\begin{equation}
\alpha=\sin^{-1}[(1+z)\sin\theta]-\theta
\end{equation}
\begin{equation}
\frac{d\alpha}{d\theta}=\frac{(1+z)\cos\theta}{\sqrt{1-(1+z)^2\sin^2\theta}}-1.
\end{equation}
The derivative has a singularity at
\begin{equation}
\theta_h=\sin^{-1}\frac{1}{1+z},
\end{equation}
which corresponds to the horizon angle of eq. [\ref{horizon}]. The
number density of photons at height $z$, arriving from angles in the
range $\theta$ to $\theta+d\theta$ is
\begin{equation}
dn(z,\theta)=n_\gamma(0)\,d\theta\,\frac{d\alpha}{d\theta}
\frac{\sin^2\theta}{\sin\alpha}\cos(\alpha+\theta).
\end{equation}
Suppose the protons have an energy that depends on the height over
which they have been accelerated: $\epsilon(z)$. At height $z$, the
mean-free-path $l(x,\theta)$ for $p\rightarrow\Delta^+$ conversion
through scattering with a photon 
arriving at an angle between $\theta$ and $\theta+d\theta$ is
given by
\begin{equation}
l^{-1}(z,\theta)=dn(z,\theta)\,\sigma(\epsilon(z),\theta).
\end{equation}
To obtain the total mean-free-path for conversion at height $z$, 
we integrate over all possible angles for the incident photons:
\begin{equation}
l^{-1}(z)=\int_{\theta_0}^{\theta_h}d\theta
\frac{dn(z,\theta)}{d\theta}\sigma(\epsilon(z),\theta).
\end{equation}
The probability for $p\rightarrow\Delta^+$ conversion of a proton as it 
moves from $z$ to $z+dz$ is
\begin{equation}
dP_c=\frac{R}{l(z)} dz.
\end{equation}
Let us now assume a specific dependence of the proton energy on
height:
\begin{equation}
x=\left(\frac{z}{L}\right )^\gamma,
\end{equation}
where $L$ is the characteristic acceleration length of the
proton in units of $R$. Then
\begin{equation}
dz = \frac{L}{\gamma}x^{1/\gamma-1}dx.
\end{equation}
The probability of conversion per unit (dimensionless) energy
interval is
\begin{equation}
\frac{dP_c}{dx}=\frac{R Lx^{1/\gamma-1}}{\gamma}
\int_{\theta_0}^{\theta_h}d\theta
\frac{dn(x,\theta)}{d\theta}\sigma(x,\theta)
\label{dpdx}
\end{equation}
This is the energy spectrum of converted $\Delta^+$ particles. The neutrino
spectrum is the same, but is scaled down in energy according
to $x_{\nu}=0.05x$. The amplitude increases by a factor of 20 to
maintain total probability. From eq. (\ref{diffflux}) we find (in
units of {GeV$^{-1}$ m$^{-2}$ s$^{-1}$)
\begin{equation}
\frac{d\phi_\nu}{d\epsilon_\nu}=1.1\times 10^{-3} f_b 
B_{12} p_{\rm ms}^{-1} d_{\rm kpc}^{-2} \frac{dP_c}{d\epsilon_\nu}
\end{equation}
where $n_0=7\times 10^{13} B_{12} 
p_{\rm ms}^{-1}$ $\rm cm^{-3}$. We obtain 
$dP_c/d\epsilon_\nu$ from eq. (\ref{dpdx}). Since $x_\nu=0.05 x$, we
have
\begin{equation}
\frac{dP_c}{d\epsilon_\nu}=20\frac{\epsilon_\gamma}{\epsilon_0^2}\frac{dP_c}{dx}
= 
2.7\times 10^{-5}\, T_{\rm 0.1
keV}\frac{dP_c}{dx} \,\, \mbox{GeV$^{-1}$}.
\label{dpde}
\end{equation}
Combining eq. (\ref{dpde}) with eq. (\ref{diffflux}) gives 
the neutrino energy flux:
\begin{equation}
\frac{d\phi_\nu}{d\epsilon_\nu}=3 \times 10^{-8} f_b 
B_{12} p_{\rm ms}^{-1} d_{\rm kpc}^{-2} T_{\rm 0.1 keV}
\frac{dP_c}{dx} 
\label{dphide}
\end{equation}
Eq. (\ref{dphide}) is our chief result. It allows comparison with
observed event distributions as well as an estimate of the total
expected counts. In Fig. \ref{spectrum} we show the neutrino energy
flux for two candidate sources, the Crab and Vela pulsars, located
respectively in the northern and southern hemispheres. For
illustration, we consider linear ($\gamma=1$) and a quadratic
($\gamma=2$) proton acceleration laws. Linear acceleration corresponds
to an accelerating field that is constant in space. Quadratic
acceleration corresponds to an accelerating field that grows linearly
with height above the star. We have used $x_w=0.1$.

For either acceleration law, the spectrum
begins sharply at
\begin{equation}
\epsilon_\nu \simeq 70 T_{\rm 0.1keV}^{-1}\mbox{ TeV}, 
\end{equation}
corresponding to the onset of the resonance. At higher energies, the
spectrum drops approximately as $\epsilon_\nu^{-2}$, as the phase space for
conversion becomes restricted; higher energy neutrinos are produced by
protons that have been accelerated to greater heights, where the
photon density is lower and the solid angle subtended by the star (as seen
by the proton) is smaller. At some maximum energy, the spectrum is
suddenly truncated by either kinematics (solid curve) or the
termination of the proton acceleration as limited by the magnitude of
the acceleration gap (not shown, since this cut-off has not been
predicted). 

In obtaining eq. (\ref{dphide}), we neglected the effects of general
relativity, except when relating the stellar temperature at
infinity to the temperature near the surface. General relativistic
effects bend the photon trajectories and bring some photons from
beyond the classical stellar horizon (eq. \ref{horizon}) to where they
can interact with the protons. This effect
enhances the rate, but not significantly. To estimate the magnitude of
this effect, we can regard the star as effectively larger, and replace
the stellar radius that appears as a pre-factor in eq. (\ref{dpdx})
by the effective stellar radius at infinity, $(1+z_g)R$. The effects
of gravitational light bending will increase the flux by a factor of
$\simeq 1+z_g\simeq 1.4$ for $R=9$ km and a factor of only $1.2$ for
$R=12$ km. In light of the large uncertainties in the values of 
$f_d$ and $L$, we simply took $R=10$ km in eq. (\ref{dpdx}) for
the purpose of making estimates.

\begin{table*}
\begin{tabular}{lllllllll}
Source & $d_{\rm kpc}$ & age & $p_{\rm ms}$ & $B_{12}$ & $T_{\rm
0.1keV}$
& $f_b$ & $P_c$ & $dN/dAdt$ \\
 & & yr & & & &  & & km$^{-2}$ yr$^{-1}$   \\
\hline
Crab & 2 & $10^3$ & 33 & 3.8 & $\le 1.7$ (Weisskopf et al., 2004) & 0.14 & $\rm 1.6\times 10^{-3}$ & 45 \\
Vela & 0.29 & $10^{4.2}$ & 89 & 3.4 & 0.6 (Pavlov et al., 2001)& 0.04 & $\rm 7.2\times 10^{-5}$ & 25 \\
J0205+64 & 3.2 & $10^{2.9}$ & 65 & 3.8 & $\le 0.9$ (Slane et al., 2002)& 0.05 & $\rm 2.4\times 10^{-4}$ & 1 \\
B1509-58 & 4.4 & $10^{3.2}$ & 151 & 15 & 1? & 0.26 & $\rm 3.4\times 10^{-4}$ & 5 \\
B1706-44 & 1.8 & $10^{4.3}$ & 102 & 3.1 & 1? & 0.13 & $\rm 3.4\times 10^{-4}$ & 5 \\
B1823-13 & 4.1 & $10^{4.3}$ & 101 & 2.8 & 1? & 0.34 & $\rm 3.4\times 10^{-4}$ & 2 \\
Cass A & 3.5 & 300 & 10? & 1? & 4 (Pavlov et al., 2004)& 0.1? &
$2.1\times 10^{-2}$ & 50 \\
SN 1987a & 50 & 17 & 1? & 1? & 4? & 0.1? & $2.1\times 10^{-2}$ &  3 \\
\end{tabular}
\caption{Estimated upper limits on the $\mu$ fluxes at Earth. 
Numbers followed by
question marks indicate guesses. Radio pulsar spin parameters were taken from the catalogue of the
Parkes Radiopulsar Survey
(Manchester et al. 2005). Temperatures and limits
on temperatures were taken from the references indicated. The
temperature upper limits on the Crab and J0205+64 were used. The
integrated conversion probability $P_c$ is reported for the case of
linear acceleration, and assuming $L=0.1$.}
\end{table*}

\section[]{Estimated Count Rates}

We now use the spectrum obtained in the previous section to estimate
the count rate in a detector.  Large-area neutrino detectors use the
Earth as a medium for conversion of a muon neutrino to a muon, which
then produces
\v{C}erenkov light in the detector. The conversion probability in
the Earth is (Gaisser et al. 1995): 
\begin{equation}
P_{\nu_\mu \rightarrow \mu} \simeq 1.3 \times 10^{-6}
\left(\frac{\epsilon_\nu}{\mbox{1 TeV}}\right)
\end{equation}
The muon event rate is
\begin{equation}
\frac{dN}{dAdt} = \int_{\epsilon_T}^{10\epsilon_T} d\epsilon_\nu \frac{d\phi_\nu}{d\epsilon_\nu}
P_{\nu_\mu \rightarrow \mu}. \label{rate}
\end{equation}
The choice of the upper limit of integration is not very important,
because the spectrum is steep. Estimated count rates are given in the
last column of Table 1 for the Crab, Vela and 9 other pulsars,
assuming $L=0.1$ and linear acceleration.  Estimated conversion
probabilities for $p\rightarrow\Delta^+$, obtained by integrating
eq. (\ref{dpdx}) from $x_T$ to $10x_T$, are shown in the penultimate
column. Our event rates are a factor of $\sim 10-30$ lower than
estimated in LB05. The main reason for the lowered rate is phase space
limitations imposed by the geometry; protons at a given height can
only undergo resonant conversion from photons arriving from the
surface in a narrow range of angles. Moreover, there is a competition
between the finite distance over which protons reach sufficient
energy to be converted, and the reduction in the number of photons
arriving from the star with the correct angular range for resonant
conversion to be kinematically allowed.

\begin{figure}
\centering
 \includegraphics[width=6cm, angle=270]{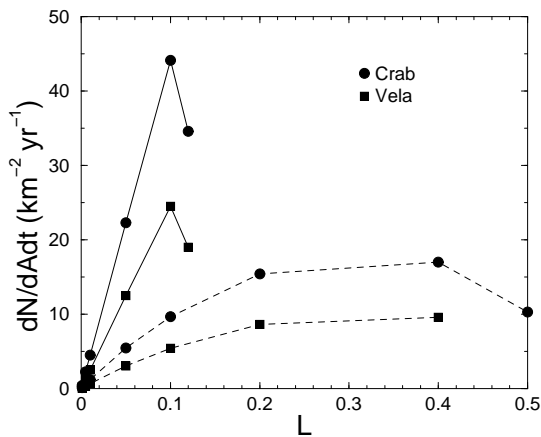}
  \caption{The muon event rate is shown as a function of the acceleration length
  for the Crab (circles) and the Vela (squares) pulsars. Upper (lower) curves
  display results for the case of linear (quadratic) proton
  acceleration. Solid and dashed lines guide the eye. }\label{event_rate}
\end{figure}

In Fig. \ref{event_rate} we show the muon event rates estimated for
the Crab (circles) and the Vela pulsars (squares), as a function of
the acceleration length $L$. The upper curves refer to calculations
performed for a linear acceleration law, whereas the lower ones assume
a quadratic one. The event rate increases up to a maximum
value, and then decreases. Increasing $L$ initially enhances the rate
because the protons then attain enough energy to undergo resonant
scattering at a height where much of the stellar surface is visible. If
$L$ is made too large, however, the process is reduced by a lower
photon density and restricted range in photon angles available for
scattering. The dependence on $L$ is more gradual for the quadratic
acceleration model, because the protons must travel farther to attain
enough energy for resonance. 

\section[]{Discussion}

To summarize, if protons reach the photomeson production resonance in
the neutron star magnetosphere, they will produce a spectrum of muon
neutrinos with the following simple characteristics:
\begin{enumerate}
\item A sharp turn on at 
\begin{equation}
\epsilon_\nu \simeq 70 T_{\rm 0.1keV}^{-1}\mbox{ TeV}, 
\end{equation}
corresponding to the onset of the resonance. 
\item A rapid fall with energy as $\epsilon_\nu^{-2}$, determined by
scattering kinematics. 
\end{enumerate}

Neutrinos are produced at relatively high rates only if the protons
are accelerated through the resonance close to the star ($L\lap 1$;
see Fig. 3). We obtain integrated count rates of several to $\sim 100$
km$^{-2}$ yr$^{-1}$ for a depletion factor $f_d\simeq 1/2$. Such count
rates should be easily detected by IceCube, and possibly by AMANDA-II
or ANTARES with integration times of about a decade (IceCube is
planned to have replaced AMANDA-II by then). While the characteristics
of the spectrum presented here are robust, we caution that the event
rates we obtain are very rough upper limits, subject to many
uncertainties. For example, we have assumed that the neutrinos are
beamed into the same solid angle as the radio beam, which might not be
a correct assumption. The radio beam is thought to be produced within
$\sim 10R$ (see, e.g., Cordes 1978). In our model, the pions are
produced much closer to the star. They then propagate to $\sim 1000R$
before decaying to neutrinos. At this distance from the star, the
field is not dipolar, and it is difficult to say anything definite
about the distribution of pion trajectories in this region. If the
neutrinos form a beam, it may be more or less collimated than the
radio beam. If the neutrino beam is more collimated, the neutrino
event rates would be higher than estimated here.

The first 807 d of data from AMANDA-II revealed no statistically
significant sources (Gro\ss, 2005). Intriguingly, there were 10 events
(over a background of 5.4) recorded from the direction of the Crab
pulsar; IceCube will be able to confirm or refute this result.  While
it would be more exciting to actually see neutrinos from pulsars, the
accumulation of null results over the next decade would be interesting
as well; it would probably mean that photomeson production is
ineffective or non-existent in the neutron star magnetosphere, thus
providing a bound on the accelerating potential.

\section*{Acknowledgments}
B.L. thanks both the INFN Sezione di Catania and the University of
Pisa for their hospitality, where much of this work was performed.

\bsp

\label{lastpage}

\end{document}